\documentclass[prl,superscriptaddress,twocolumn]{revtex4}
\usepackage{dcolumn}
\usepackage{bm}
\usepackage{amssymb}
\usepackage{amsmath}
\usepackage{amsfonts}
\usepackage[english]{babel}
\usepackage{amsmath}
\usepackage{graphicx}
\usepackage{dcolumn}
\usepackage{bm}
\usepackage{hyperref}
\usepackage{bigints}

\renewcommand{\d}{\mathrm{d}}

\RequirePackage[cdot,textstyle,amssymb]{SIunits}%
\RequirePackage{units}
\usepackage[usenames, dvipsnames]{color}

\usepackage[]{graphicx}

\usepackage{epstopdf}
\begin{document}

\title{Micromagnetics of anti-skyrmions in ultrathin films }
\author{Lorenzo Camosi} \email{lorenzo.camosi@neel.cnrs.fr}
\affiliation{Univ.~Grenoble Alpes, CNRS, Institut N\'eel, F-38000 Grenoble, France}
\author{Nicolas Rougemaille}
\affiliation{Univ.~Grenoble Alpes, CNRS, Institut N\'eel, F-38000 Grenoble, France}
\author{Olivier Fruchart}
\affiliation{Univ.~Grenoble Alpes, CNRS, CEA, Grenoble INP, INAC-SPINTEC, F-38000 Grenoble, France}
\author{Jan Vogel} \email{jan.vogel@neel.cnrs.fr}
\affiliation{Univ.~Grenoble Alpes, CNRS, Institut N\'eel, F-38000 Grenoble, France}
\author{Stanislas Rohart}
\affiliation{Laboratoire de Physique des Solides, Universit\'{e} Paris-Sud, Universit\'{e} Paris-Saclay, CNRS UMR 8502, F-91405 Orsay cedex, France}

\begin{abstract}
We present a combined analytical and numerical micromagnetic study of the equilibrium energy, size and shape of anti-skyrmionic magnetic configurations. Anti-skyrmions can be stabilized when the Dzyaloshinskii-Moriya interaction has opposite signs along two orthogonal in-plane directions, breaking the magnetic circular symmetry. We compare the equilibrium energy, size and shape of anti-skyrmions and skyrmions that are stabilized respectively in environments with anisotropic and isotropic Dzyaloshinskii-Moriya interaction, but with the same strength of the magnetic interactions.
When the dipolar interactions are neglected the skyrmion and the anti-skyrmion have the same energy, shape and size in their respective environment. However, when dipolar interactions are considered, the energy of the anti-skyrmion is strongly reduced and its equilibrium size increased with respect to the skyrmion. While the skyrmion configuration shows homochiral N\'{e}el magnetization rotations, anti-skyrmions show partly N\'{e}el and partly Bloch rotations. The latter do not produce magnetic charges and thus cost less dipolar energy. Both magnetic configurations are stable when the magnetic energies almost cancel each other, which means that a small variation of one parameter can drastically change their configuration, size and energy.
\end{abstract}

\maketitle

\section{Introduction}

The prediction \cite{Bogdanov1989,Ivanov1990} and first experimental observations \cite{Muhlbauer2009,Yu2010,Heinze2011,Jiang2015,Boulle2016} of skyrmions has led to a huge increase of research on materials suitable for hosting this kind of structures. Skyrmions are chiral magnetic solitons, which can be stabilized by a chiral interaction like the Dzyaloshinskii-Moriya interaction (DMI) \cite{Dzyaloshinskii1957,Moriya1960} or by dipolar interactions. DMI is an anti-symmetric exchange interaction that favors perpendicular alignment between neighboring moments, which may exist in systems that lack a structural inversion symmetry.
The presence of anisotropic DMI with opposite signs along two perpendicular directions can stabilize anti-skyrmions \cite{Camosi2017,HoffmannM2017}. Skyrmions (Sk) and anti-skyrmions (ASk) are characterized by their topological charge $N_{\mathrm{sk}}$ \cite{Nagaosa2013}. It yields some of their most fascinating properties, such as the topological Hall effect \cite{Neubauer2009,Li2013} or the topological gyroscopic effect, also called the skyrmion Hall effect \cite{Zang2011,Jiang2017,Klaui2011}.
In a continuous-field approximation $N_{\mathrm{sk}}$ can be written as the integral on the space $(r ,\alpha)$ that counts how many times the magnetization $\mathbf{m}$ wraps the unit sphere \cite{Nagaosa2013}.

\begin{figure}[h]
  \begin{center}
    \includegraphics[width=9cm]{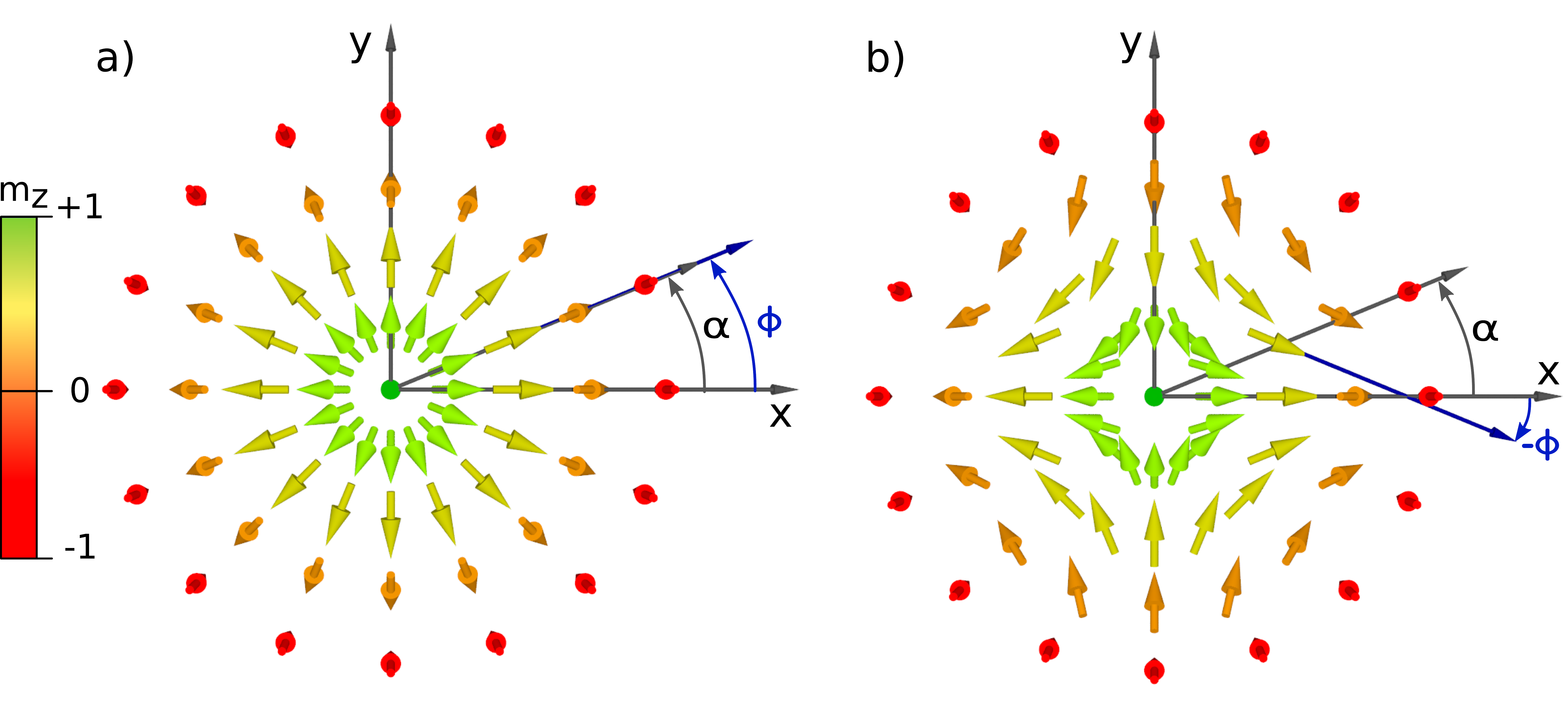}
  \end{center}
  \caption{\label{fig:SKASK} Sketch of the skyrmion and anti-skyrmion configurations. The illustration of  $\alpha$, the space angle in the plane, and $\phi$, the in-plane magnetization angle, show the difference between a skyrmion ($\frac{d \phi}{d \alpha} = 1$) and an anti-skyrmion ($\frac{d \phi}{d \alpha} = -1$).}
\end{figure}

\small
\begin{equation}
\label{eq:topo}
N_{\mathrm{sk}}= \frac{1}{4\pi} \iint \frac{d \theta}{d r} \frac{d \phi}{d \alpha} \sin \theta \,dr \,d\alpha  = p W = \pm 1
\end{equation}
\normalsize

\noindent where $\theta$ and $\phi$ are the polar and azimuthal coordinates of $\mathbf{m}$ (Fig.~\ref{fig:SKASK} and \ref{fig:frame}), \textit{p} describes the direction of magnetization in the core of the texture [$p = \cos\theta(r=0)$, with $\theta(r=0)$ = 0 or $\pi$] and $ W =[ \phi(\alpha) ]^{\alpha=2 \pi}_{\alpha= 0}/2 \pi= \pm 1 $ is the winding number. Both Sk and ASk can thus have topological charge $+1$ or $-1$, but they have opposite winding numbers ($\frac{d \phi}{d \alpha} = 1$ for a Sk and $\frac{d \phi}{d \alpha} = -1$ for an ASk) and hence opposite topological charges for a given direction of the core magnetization \textit{p}.

Sk have been observed both in bulk materials and thin films \cite{Muhlbauer2009,Yu2010,Heinze2011,Jiang2015,Boulle2016}.
ASk have been observed in bulk materials \cite{Nayak2017}, but not yet in thin film systems. The reason is that most thin film systems showing DMI studied until now were polycrystalline, leading to the same sign and strength of the DMI along any in-plane direction. In order to stabilize ASk in thin films with perpendicular magnetization, the sign of DMI has to be opposite along two in-plane directions of the film. This may occur in epitaxial thin films with a $C_{2v}$ symmetry \cite{Camosi2017,Gungordu2016,HoffmannM2017}. For a Au/Co/W(110) thin film with $C_{2v}$ symmetry, we have recently shown experimentally that the crystal symmetry can indeed give rise to an anisotropic DMI \cite{Camosi2017}. In this system, the DMI is 2.5 times larger along the \textrm{bcc}$[\overline{1}10]$ direction than along the \textrm{bcc}$[001]$ direction, but has the same sign. It was shown theoretically that an opposite sign of the DMI, needed for ASk, may be found for instance in a Fe/W(110) thin film \cite{HoffmannM2017}.\\

The Sk and ASk dynamics has already been theoretically investigated. If we consider the dynamics driven by spin-polarized currents, we can distinguish two main effects. The first is due to the different torques acting on the moment configurations, leading to a motion which keeps the same angle with respect to the current direction for Sk, while it is strongly anisotropic for ASk \cite{Huang2017}.
The second is given by the topological gyroscopic effect (skyrmion Hall effect) \cite{Thiele1973,Jiang2017,Klaui2011}. In the same ferromagnetic state, the two magnetic configurations have opposite topological numbers, leading to opposite lateral deviations \citep{Kiselev2017,Everschor2017}. The combination of the two effects leads to the suppression or enhancement of the ASk transverse motion depending on the current directions \cite{Huang2017}. Moreover, the coupling of a Sk and an ASk may lead to the absence of a skyrmion Hall effect \cite{Huang2017}.\\

A numerical micromagnetic study of the stability of Sk and ASk in frustrated ferromagnets has been reported by Xichao et al. \citep{Xichao2017}. They evidenced the role of the dipolar interaction in the ASk stabilization. Our paper presents a complete combined analytical and numerical micromagnetic study of the equilibrium energy, size and shape of ASk in ferromagnetic thin films where the DMI has the same strength but opposite signs along two orthogonal in-plane directions. We point out the quantitative differences with the Sk magnetic configuration.
The paper is divided into two parts. In the first part we show that the micromagnetic energy of an ASk can be written in a circular symmetric form and that neglecting the dipolar interaction the ASk and the Sk have the same equilibrium size, shape and energy. In the second part we investigate the influence of the difference in dipolar energy on both the stability and the equilibrium shapes and sizes of Sk and ASk.

\section{Anti-skyrmions in the absence of long-range dipolar interactions}

It has been shown that skyrmionic configurations can be stabilized by DMI and/or dipolar interactions (in thin films) (\cite{Bogdanov2001,Rossler2006,Leonov2016,Rohart2013}). For a small Sk, the long-range dipolar interaction can be neglected \cite{Anne2017}. Sks are then stabilized only by the competition between DMI, exchange and magnetic anisotropy \citep{Bogdanov1989} which we consider in this part. We develop micromagnetic calculations in order to study the energy and the size of Sk and ASk configurations.

Using the notation of the Lifshitz invariants $L^{(i)}_{jk}= m_j \frac{\partial m_k}{\partial i}-m_k \frac{\partial m_j}{\partial i} $, the total micromagnetic DM energy can be written as :

\begin{equation}
E_\mathrm{DMI} = t \iint \left(D_x L_{xz}^{(x)}+D_yL_{yz}^{(y)}\right) \d x\d y
\end{equation}

\noindent where D$_x$ and D$_y$ are the DMI strengths along the $\mathbf{x}$ and $\mathbf{y}$ directions, $m_x$, $m_y$ and $m_z$ are the components of the unit magnetization vector $\mathbf{m}$ and $t$ is the film thickness. We focus on the simplest case where the absolute value of the DMI is equal along the principal directions $\mathbf{x}$ and $\mathbf{y}$, with the same sign for skyrmions ($D_x=D_y$) but opposite in sign for anti-skyrmions ($D_x=-D_y$). The condition $D_x = -D_y$ can be found, for example, in D2d and S4 crystal symmetry classes~\cite{Bogdanov1989}. In that case, the micromagnetic DM energy for an anti-skyrmion can be written as :

\begin{equation}
E_\mathrm{DMI} = t \iint D\left(L_{xz}^{(x)}-L_{yz}^{(y)}\right) \d x\d y
\end{equation}

\noindent where $D = D_x$. For simplicity, the exchange $E_{\mathrm{ex}}$ and magnetic anisotropy $E_{K}$ energies are considered to be isotropic in the thin film plane $(\mathbf{x},\mathbf{y})$ :

\begin{equation}
E_\mathrm{ex} = t \iint A\left(\nabla\mathbf{m}\right)^2\d x\d y
\end{equation}

\noindent with $A$ the exchange stiffness and $\left(\nabla\mathbf{m}\right)^2=(\partial \mathbf{m}/\partial x)^2+(\partial \mathbf{m}/\partial y)^2$, and

\begin{equation}
E_\mathrm{K} = -t \iint K_\mathrm{eff}(\mathbf{m}.\mathbf{e_z})^2\d x\d y
\label{eq:K}
\end{equation}

\noindent The only dipolar interaction taken into account is the one due to magnetic surface charges, which in a local approximation can be treated as an effective anisotropy : $K_\mathrm{eff}=K_\mathrm{0}-\frac12\mu_0M_\mathrm{s}^2$, where $K_\mathrm{0}$ is the sum of the magneto-crystalline and interface anisotropies. Here we consider a system with perpendicular anisotropy, therefore $K_\mathrm{eff} > 0$. The anisotropy and the exchange do not depend on the in-plane magnetization, and therefore they are the same for a Sk and an ASk of the same size.\\

The formulation of the complete micromagnetic energy for an arbitrary magnetization texture involves a large number of degrees of freedom. They can be strongly reduced using circular symmetry as it was done to solve the Sk profile \cite{Bogdanov2001}. However the circular symmetry is broken for an ASk by the presence of the chiral inversion. To approach this problem, we first consider a 1D modulation, propagating in an arbitrary direction in the plane. Then we use the obtained result to solve the ASk 2D profile.

    \subsection{1D modulation}

\begin{figure}[h]
  \begin{center}
    \includegraphics[width=8cm]{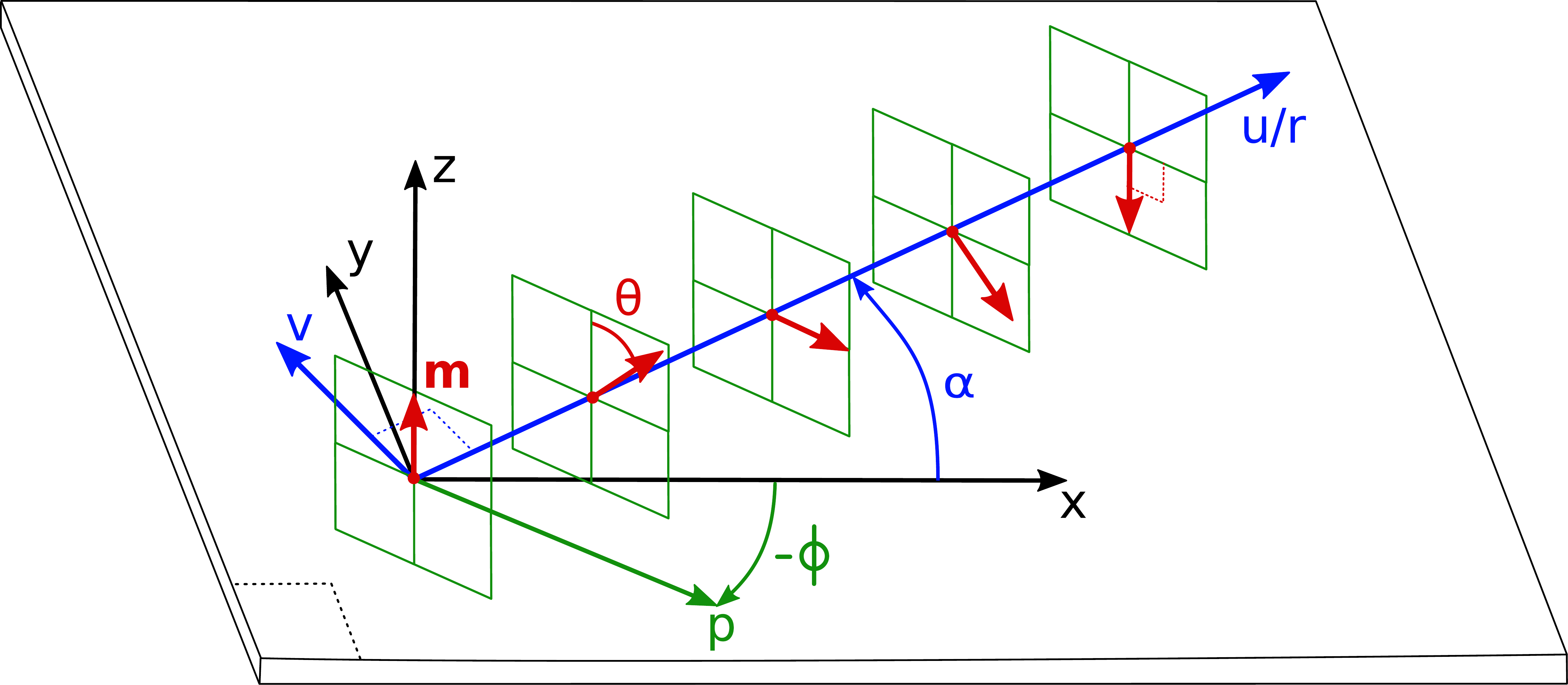}
  \end{center}
  \caption{\label{fig:frame} Frames used to describe a 1D modulation and 2D skyrmionic texture in a C$_{2v}$ system }
\end{figure}

In the general case of isotropic DMI ($D_x= D_y$) the energy is an invariant upon in-plane framework rotation. Here, we consider a $C_{2v}$ symmetry plane, where the DMI has opposite chirality along perpendicular directions ($D_x= -D_y$).
In this system we introduce a 1D modulation propagating in a direction $\mathbf{u}$ oriented at an angle $\alpha$ with respect to the $\mathbf{x}$ axis (Fig.~\ref{fig:frame}).

Indeed in the rotated framework $(\mathbf{u},\mathbf{v})$, the integrated Dzyaloshinskii-Moriya interaction for the 1D modulation reads:

    \begin{equation}
    E_\mathrm{DMI} = t \iint D\left[ \cos(2\alpha) L_{uz}^{(u)}-\sin(2\alpha) L_{vz}^{(u)}\right]\d u \d v
    \end{equation}

\noindent This form promotes N\'eel type modulations with opposite chiralities for $\alpha = 0$ or $\pi/2$, and chiral Bloch type modulations for $\alpha = \pi/4$ and $3\pi/4$ \citep{Camosi2017}.
Along the modulation, the magnetization $\mathbf{m}$ lies in the $(\mathbf{p},\mathbf{z})$ plane, where $\mathbf{p}$ represents the modulation polarization (Fig.~\ref{fig:frame}). It is therefore characterized by $\theta(u)$, the angle between $\mathbf{m}(u)$ and the $\mathbf{z}$ axis, and $\phi$, the angle between $\mathbf{p}$ and the $\mathbf{x}$ axis.
The micromagnetic energy per unit width becomes :

    \small
    \begin{equation}
        w = t \int \left[ A\left(\frac{d\theta}{du}\right)^2 - D\cos(\alpha+\phi)\frac{d\theta}{du} + K\sin^2\theta \right]\d u
    \end{equation}
    \normalsize

\noindent Notice that the first (Heisenberg exchange, with exchange stiffness $A$) and third (effective magnetic anisotropy) terms do not depend on the propagation direction $\mathbf{u}$ nor on the polarization $\mathbf{p}$. On the contrary, the local density of Dzyaloshinskii-Moriya energy is strongly affected. Hence the polarization direction can be determined by minimizing the DMI term $-D\cos(\alpha+\phi)$. Therefore, for positive (resp. negative) $D$, we find $\phi = - \alpha$ (resp. $\phi = - \alpha + \pi$). Such a relation exactly corresponds to the one requested for negative winding numbers (ASk). When this condition is fulfilled, the energy of the moment modulation can be formulated as a radial invariant. The micromagnetic energy per unit width of the 1D modulation becomes:

    \begin{equation}
       w = t \int \left[ A\left(\frac{d\theta}{du}\right)^2 - D\frac{d\theta}{du} + K_{\mathrm{eff}}\sin^2\theta \right] \d u
    \end{equation}

\noindent This indicates that for a given $\theta(u)$ the energy is independent of $\alpha$. Only $\mathbf{p}$ (i.e. the in-plane component of the spin modulation in Fig.~\ref{fig:frame}) needs to adapt to the anisotropic DMI. According to the specific values of $A$, $D$ and $K_\mathrm{eff}$, domain walls or spin-spirals are found with the same out-of-plane profile, as discussed in earlier papers \cite{Bogdanov2001,Rohart2013}. This means that an isotropic modulation ($\phi = \alpha$) in an isotropic environment ($D_x= D_y$) has the same energy as an anisotropic modulation ($\phi = -\alpha$) in an anisotropic environment ($D_x= -D_y$).

 \subsection{2D texture}

We extend the above calculation to a 2D texture. The texture is described by the two angles $\theta(r,\alpha)$ and $\phi(\alpha)$, defined as before, and where $r$ and $\alpha$ are the circular coordinates in the $(\mathbf{x},\mathbf{y})$ plane. The result of the 1D investigation shows that the relation $\phi = -\alpha$ is conserved. Therefore, the micromagnetic energy is isotropic, $\theta$ does not depend on $\alpha $ and the problem can be evaluated using a circular symmetry, with the total energy

    \begin{eqnarray}
   & E = 2\pi t \bigintsss \left [ A \left \lbrace \left( \frac{\d\theta}{\d u}\right)^2+\frac{\sin^2\theta}{r^2}\right\rbrace \right.\nonumber \\
         & \left.- D \left \lbrace \frac{\d\theta}{\d u} +\frac{\cos\theta\sin\theta}{r}\right\rbrace
                              + K_\mathrm{eff}\sin^2\theta \right]  r \d r \label{Eq:skyrmion_analytics}
    \end{eqnarray}

\noindent This equation is exactly the same as the one describing a Sk in a medium with isotropic DMI \citep{Bogdanov1989}. It is analogous to the calculation of Belavin and Polyakov \cite{Belavin1975}, who demonstrated that the exchange energy of a bubble does not depend on the sign of $d\phi/d\alpha$. This means that, for a given set of $A$, $D$ and $K_{\mathrm{eff}}$, the ASk texture has a profile and an energy identical to a Sk in an isotropic medium \citep{Bogdanov1989}. All the conclusions that have been drawn in preceding papers neglecting long range dipolar coupling remain valid (boundary conditions, out-of-plane profiles and energies) \cite{Bogdanov2001,Rohart2013}. The only difference between the two configurations is the $\phi(\alpha)$ relationship; $\phi = \alpha$ for a Sk ($W = 1$) and $\phi = - \alpha$ for an ASk ($W = -1$).

\begin{figure}[h]
  \begin{center}
    \includegraphics[width=9cm]{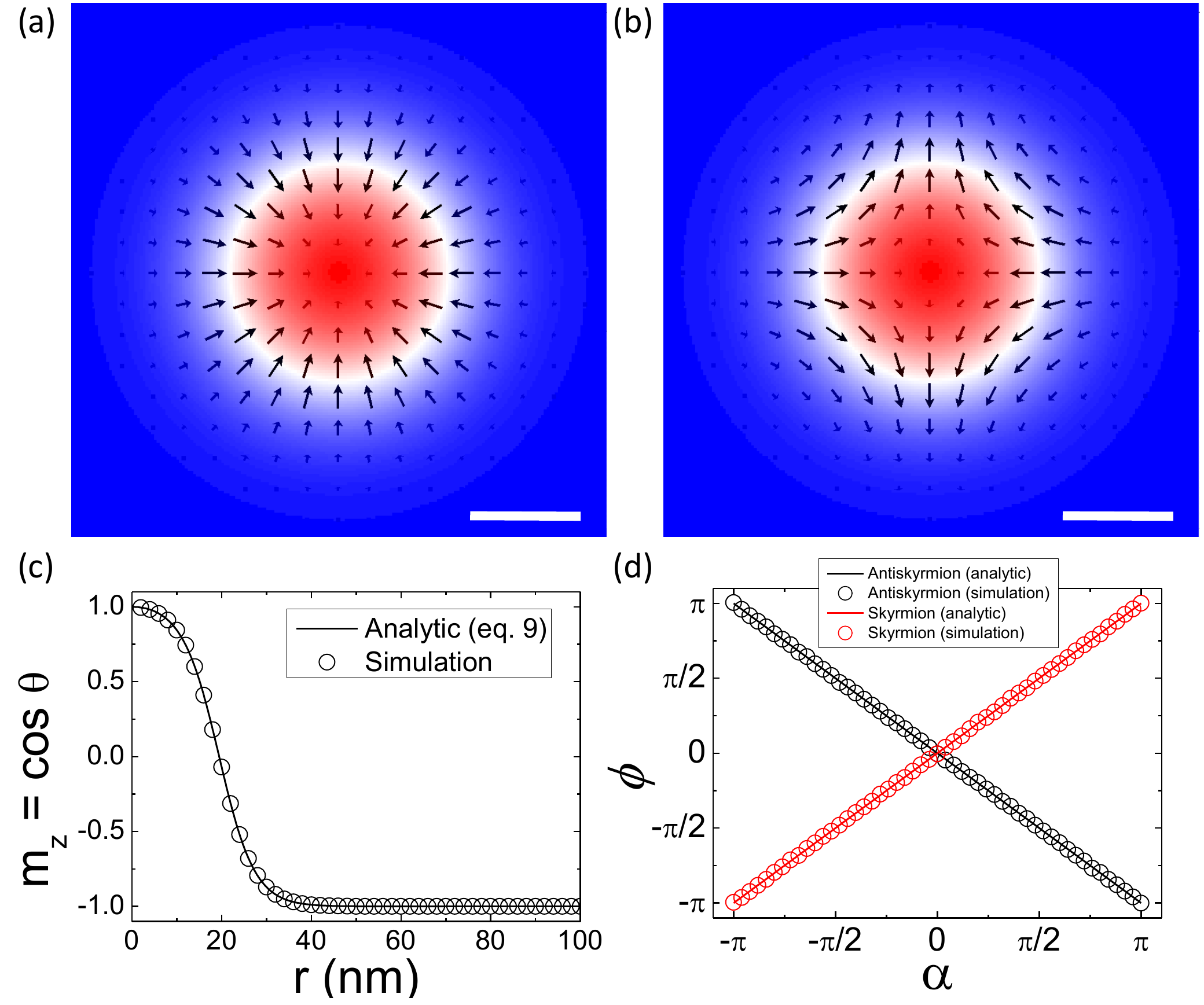}
  \end{center}
  \caption{Micromagnetic simulations of the magnetization maps for a skyrmion (a) and an anti-skyrmion (b) (scalebar : 10 nm). The arrows represent the in-plane magnetization and the color code the out-of-plane magnetization (red = up, white = in-plane and blue = down). (c) and (d) Comparison between analytic modelling and simulations of the $m_z$ profile and the $\phi(\alpha)$ relationship, for Sk and ASk. In (c) the Sk and ASk profiles are indiscernable.}\label{fig:analytics}
\end{figure}

In order to verify the validity of our analytical results we have performed micromagnetic simulations without dipolar interactions. We used an adaptation of the object-oriented micromagnetic framework code (OOMMF)~\cite{oommf, Rohart2013} including anisotropic DMI (see Fig.~\ref{fig:analytics}). The calculation is performed in a 400-nm diameter, 0.6-nm thick circular dot with typical magnetic parameters for systems where isolated skyrmions have been experimentally observed ($A=16$~pJ/m,  $K_\mathrm{eff}=0.2$~MJ/m$^3$ and $D= 2$~mJ/m$^2$) \cite{Boulle2016}. Comparing Sk and ASk obtained respectively with $D_x/D_y=1$ and $-1$, identical energies and out-of-plane profiles are found. The $\phi(\alpha)$ relationship is confirmed validating the different assumptions in our model (in particular the hypothesis that $\phi$ is independent on $r$).

\section{Role of dipolar coupling}

Determining the role of dipolar interactions on the stabilization of Sk with micromagnetic analytical calculations is particularly difficult. This interaction has often been neglected \cite{Bogdanov2001,Rossler2006,Leonov2016,Rohart2013} or analytically expressed under approximations \cite{Tu1971,Schott2017,Guslienko2015}. The two-fold symmetry of the ASk magnetic configuration does not allow using a circular symmetry, increasing the difficulty of this approach. Therefore, we performed a study of the dipolar interaction effects on the Sk and ASk configurations with the support of micromagnetic simulations using OOMMF \cite{oommf} with an anisotropic DMI. For stabilizing Sk and ASk in the absence of an external magnetic field, we confine them into 0.6\,nm thick circular dots with a diameter of 400\,nm, using a mesh size of 1\,nm.

\subsection{Phenomenology of dipolar interactions}

The effect of the dipolar interaction on the size and stability of Sk and ASk in a dot can be phenomenologically understood considering the contributions from the surface and volume charges.\\

\begin{figure}[h]
  \begin{center}
    \includegraphics[width=8cm]{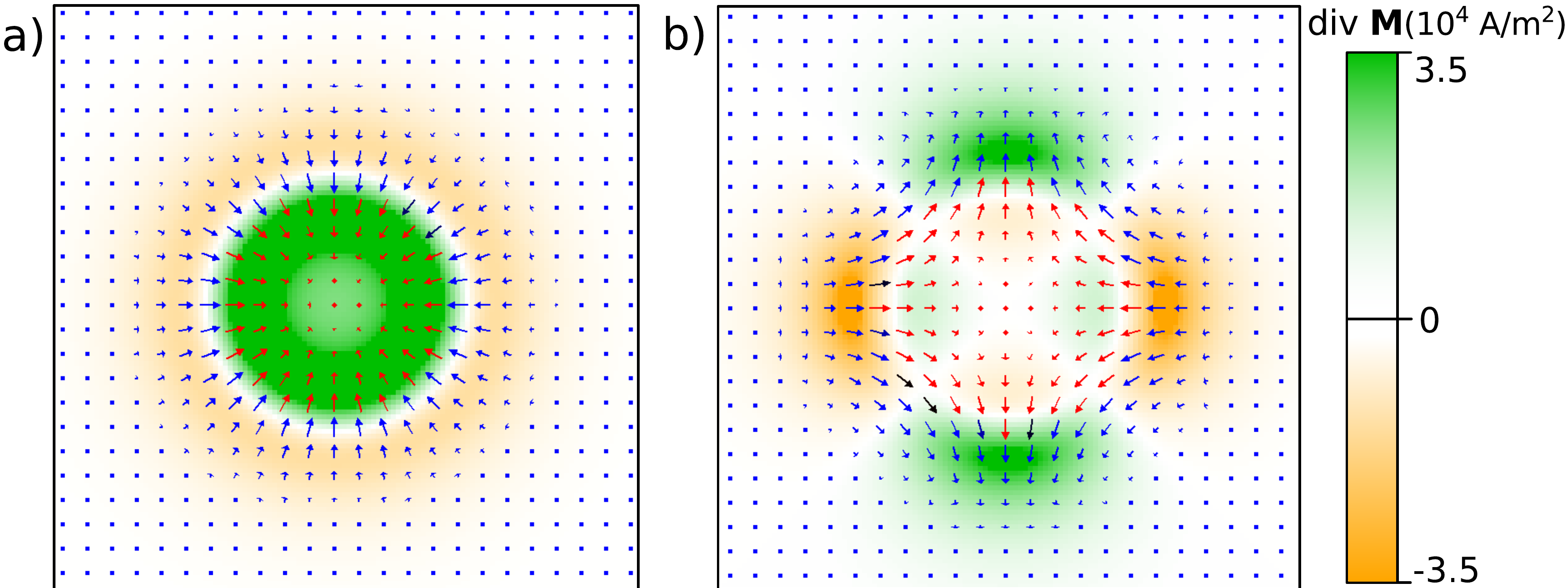}
  \end{center}
  \caption{\label{fig:volcharge} Plane view of a magnetic skyrmion \textbf{(a)} and an anti-skyrmion \textbf{(b)} in a dot with a lateral size of 115~nm. The colors represent the magnetic volume charges ($\nabla \mathbf{m}$) with the color code shown at the right. The arrows represent the magnetic configurations with their colors representing the out-of-plane component (red = up, blue = down). The radius of the magnetic configurations does not depend on dot size for dots larger than about 100nm and are stable without the application of an external magnetic field, with $M_{\mathrm{s}}= 5 \cdot 10^5\,$A/m $, A_{\mathrm{ex}}=16\, $pJ/m,  $K_{\mathrm{eff}}= 200\, $kJ/m$^3$,  $D= 2.0\, $mJ/m$^2$. Isotropic DMI ($D_x=D_y$) allows the stabilization of a skyrmion whereas anisotropic DMI ($D_x=-D_y$) stabilizes an anti-skyrmion.}
\end{figure}

Magnetic surface charges $\sigma$ arise when the magnetization vector has a component along the surface normal $\mathbf{n}$, i.e. $\sigma = M_S \mathbf{m} \cdot \mathbf{n}$. In a perpendicular thin film, they arise from the two surfaces of the film. Therefore, the texture core and the surroundings display opposite charge signs and the energy due to the dipolar interaction is reduced when the magnetic flux closes between the core and the surroudings \cite{Boulle2016,Schott2017}. Therefore a Sk or an ASk configuration confined in a dot tends to increase its radius in order to demagnetize the system \cite{Boulle2016,Schott2017}. The surface charges do not depend on the in-plane magnetization and the associated dipolar interaction is identical for a Sk and an ASk with the same shape and area.\\

Magnetic volume charges are generated from the volume magnetization divergence and are therefore present in the vicinity of magnetic textures, and strongly depend on their details. The maps of the volume charges for a Sk and an ASk configuration are shown in Fig.~\ref{fig:volcharge}. The Sk maps present a circular symmetry and the volume charges arise from the N\'eel-like magnetization rotation. The ASk maps shows a 2-fold symmetry and the presence of Bloch-like rotations along intermediate directions ($\phi=\pi/4 + n\pi/2$). For the ASk, the charges are maximum along the main axes $\mathbf{x}$ and $\mathbf{y}$, due to the N\'eel type rotations. However, the signs along $\mathbf{x}$ and $\mathbf{y}$ are opposite due to the opposite chirality.

The impact of the volume charges can be decomposed in two effects. First, the charges locally increase the energy of the transition region, in the same manner as for a linear domain wall. The dipolar energy cost of a domain wall oriented at an angle $\alpha$ with respect to the $\mathbf{x}$ axis is $\delta_N\cos^2(\phi-\alpha)$, where $\delta_N$ corresponds to the dipolar cost of a N\'eel wall~\cite{Thiaville2012}. For a large Sk or ASk, such arguments based on the energy of a linear domain wall remain qualitatively valid~\cite{Rohart2013} and lead to a significant difference between the two textures. For a skyrmion, where $\phi = \alpha$ the dipolar energy is isotropic, equal to $\delta_N$, while for an anti-skyrmion, where $\phi = -\alpha$, the energy strongly varies as a function of $\alpha$ and becomes zero in the regions where the spin rotations are Bloch-like. Considering a circular texture of radius $R$, the additional cost to the dipolar-free micromagnetic energy described earlier is $2\pi t R \delta_N$ and $\pi t R \delta_N$ for skyrmions and antiskyrmions, respectively. This means that anti-skyrmions are expected to display a lower energy and a larger equilibrium diameter than skyrmions. A second effect couples the flux arising from the volume charges over the entire texture. Inside a skyrmion, the charges have the same sign everywhere, opposite to the sign outside the texture. On the contrary, both inside and outside an anti-skyrmion the charges have signs that are opposite along the main axes. The total volume charges both inside and outside the texture thus cancel leading to a further decrease of the dipolar energy with respect to the skyrmion.\\

The total energy of the ASk is thus reduced with respect to the Sk due to the difference in the distribution of volume magnetic charges. Moreover, the presence of anisotropic volume charges may deform the ASk shape, as we will discuss later.

\begin{figure}[h]
  \begin{center}
    \includegraphics[width=6cm]{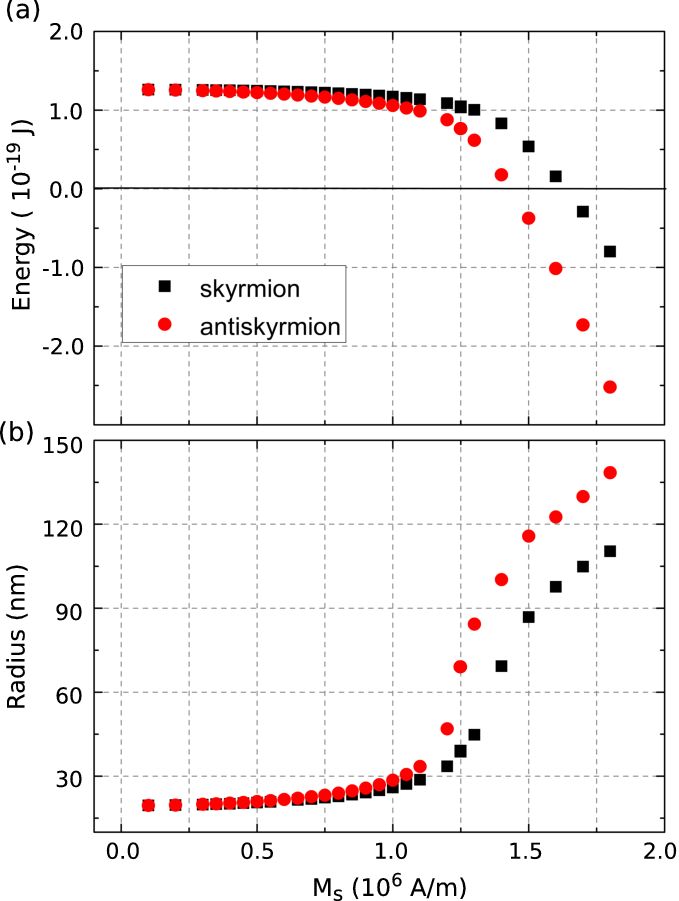}
  \end{center}
  \caption{\label{fig:E(Ms)} Energy \textbf{(a)} and radius \textbf{(b)} of a skyrmion (black) and an anti-skyrmion (red) as a function of the spontaneous magnetization $M_{\mathrm{s}}$. The simulations are performed in circular dots of 400\,nm diameter and 0.6\,nm thickness with a fixed out-of-plane effective anisotropy $K_{\mathrm{eff}}$ ($A_{\mathrm{ex}}=16\, $pJ/m, $K_{\mathrm{eff}}= 2 \cdot 10^5\, $J/m$^3$, $D= 2.0\, $mJ/m$^2$)}
\end{figure}

\subsection{Consequences for the stability of anti-skyrmions}

To investigate in more detail the effect of the dipolar interaction on the stability and shape of Sk and ASk, we studied their energy and radius as a function of the spontaneous magnetization $M_{\mathrm{s}}$ (Fig.~\ref{fig:E(Ms)}). In order to consider only the effects of the volume charges and the flux closure of the surface charges we keep $K_{\mathrm{eff}}$ constant during the variation of $M_{\mathrm{s}}$ and use the same parameters as in the simulations with simplied dipolar interactions in Fig.~\ref{fig:analytics}. $M_{\mathrm{s}}$ was varied between $0.1 \cdot 10^6\,$A/m and $1.8 \cdot 10^6 \, $A/m.
In Fig.~\ref{fig:E(Ms)}(a) the Sk and ASk energies are considered as the energy difference between a dot with a Sk or an ASk and its relative single domain state. Taking this difference allows eliminating the effect of the DMI on the edge magnetization \cite{Rohart2013,Pizzini2014}. Since the ASk can present a shape which is not circular, we consider an effective radius ($r =\sqrt{\mathcal{A}/\pi}$) calculated from the area $\mathcal{A}$. We consider$\mathcal{A}$ as the space region of the Sk and ASk where $\mathrm{m_z}>0$. For small values of $M_{\mathrm{s}}$ the Sk and the ASk are mainly stabilized by the competition between the exchange, anisotropy and DMI \cite{Anne2017} that were shown to be equal for Sk and ASk. The dipolar interaction is negligible and the Sk and the ASk have comparable energy and radius.
When $M_{\mathrm{s}}$ increases the dipolar interaction plays a larger role. The Sk and ASk radii increase (Fig.~\ref{fig:E(Ms)}(b)) to allow a more efficient flux closure between the surface magnetic charges. Both configurations lower the energy but the difference in volume charges favors the ASk. When for larger $M_{\mathrm{s}}$ the dipolar energy becomes larger than the DMI energy, the total energy of the Sk and the ASk decreases and their radius increases until they feel the repulsive effect from the dot edge \cite{Rohart2013}. In this regime, the Sk and ASk shape and dimensions strongly depend on the symmetry and size of the microstructures in which they are confined and the volume charges become the driving force for defining the magnetic configuration.

\begin{figure}[h]
  \begin{center}
    \includegraphics[width=7cm]{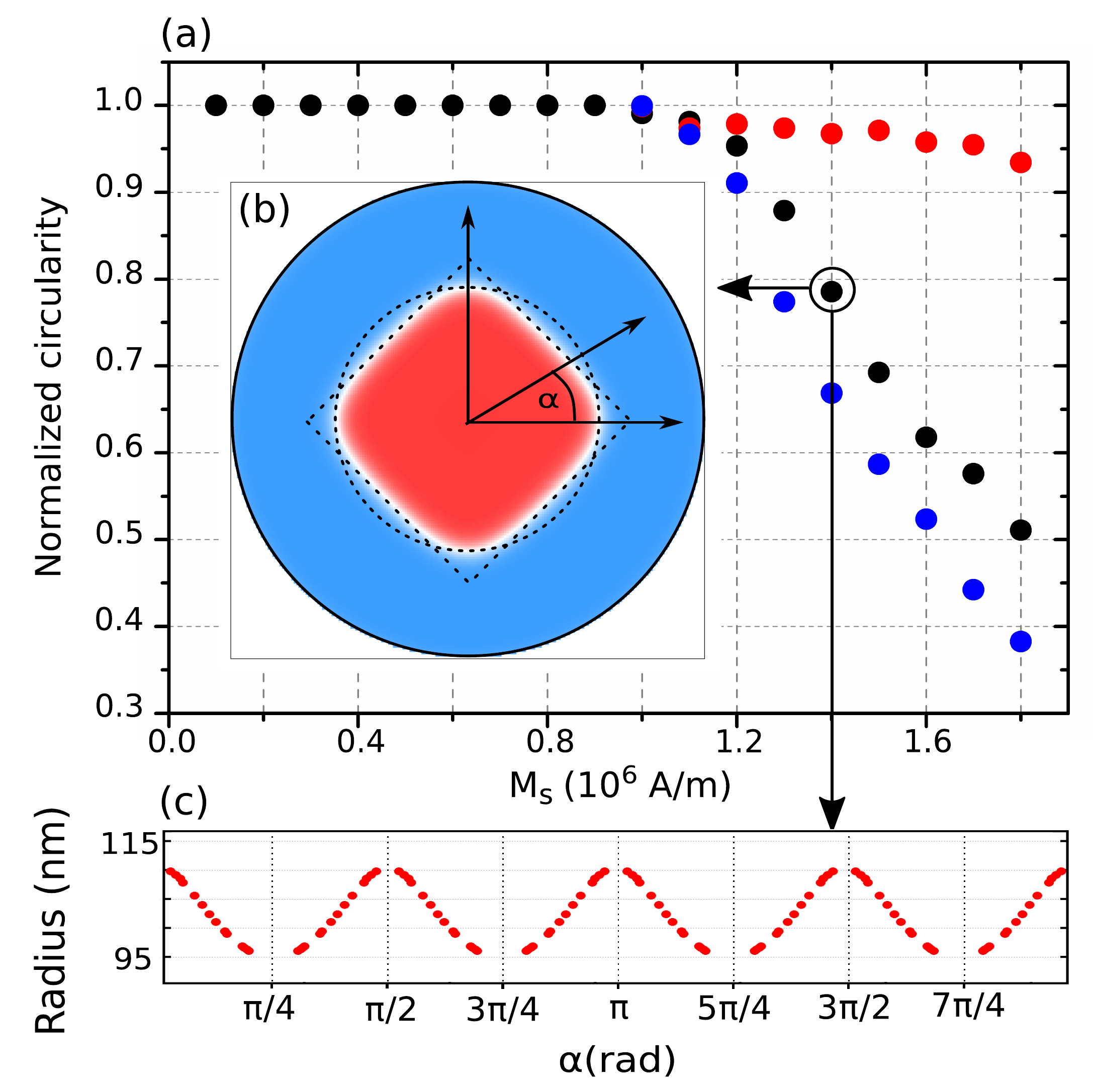}
  \end{center}
  \caption{\label{fig:circ} \textbf{(a)} Normalized circularity factor $C_n$ as a function of $M_{\mathrm{s}}$ for an anti-skyrmion stabilized in dots with a diameter of 500 nm (blue), 400 nm (black) and 200 (red). \textbf{(b)} Magnetic configuration of an anti-skyrmion ($M_{\mathrm{s}}= 1.4 \cdot 10^6\,$A/m,  $ A_{\mathrm{ex}}=16\, $pJ/m,  $K_{\mathrm{eff}}= 2 \cdot 10^5\, $J/m$^3$,  $D= 2.0\, $mJ/m$^2$) with a sketch that shows the square-circular shape. \textbf{(c)} Radius of an anti-skyrmion as a function of the in-plane angle $\alpha$ }
\end{figure}

Upon increasing $M_{\mathrm{s}}$, the ASk changes its shape in order to promote Bloch-like rotations, which do not generate volume charges and thus cost less dipolar energy. Because DMI promotes Bloch-like rotations along intermediate crystallographic directions ($\phi=\pi/4 + n\pi/2$) the ASk has the tendency to acquire a square shape (Fig.~\ref{fig:circ}) \cite{discret}. This configuration allows increasing the ratio between Bloch and N\'eel rotations. Even if the total domain wall length increases going from a circular to a square shape, this can still lead to a minimisation of the total DMI and dipolar energy.

In order to quantify this tendency, we calculate the normalized circularity factor $C_n = \frac{4 \pi \mathcal{A} / \mathcal{P}^2 - \pi/4}{ 1 - \pi/4}$, where $\mathcal{A}$ and $\mathcal{P}$ are, respectively, the area and the perimeter of the texture (set of points where $\mathrm{m_z}=0$). This normalized circularity factor may vary from $C = 1$ (circle) to  $C = 0$ (square). The Sk has a circular symmetry and this factor is thus equal to 1~\cite{discret}. Fig.~\ref{fig:circ} shows the plot of the normalized circularity for an ASk as a function of $M_{\mathrm{s}}$, stabilized in dots of different diameters (200, 400 and 500 nm). We can distinguish two different regimes. For small values of $M_{\mathrm{s}}$ the volume charges do not influence the ASk shape, the circularity is equal to $C = 1$ and does not depend on the dot size. For larger $M_{\mathrm{s}}$ values the moment rotation with an angle ($\phi=\pi/4 + n\pi/2$) is favored and the ASk circularity decreases upon increasing $M_{\mathrm{s}}$. In small dots, the confinement does not allow the ASk to expand, and constrains the texture shape to the dot shape. In larger dots, the ASk shape adapts to the internal energies, which results in a shape closer to the square configuration. Increasing the dot size even further, the ASk configuration becomes unstable for these large values of M$_s$ and a labyrinth-like domain structure is formed.

\begin{figure}[h]
  \begin{center}
    \includegraphics[width=8cm]{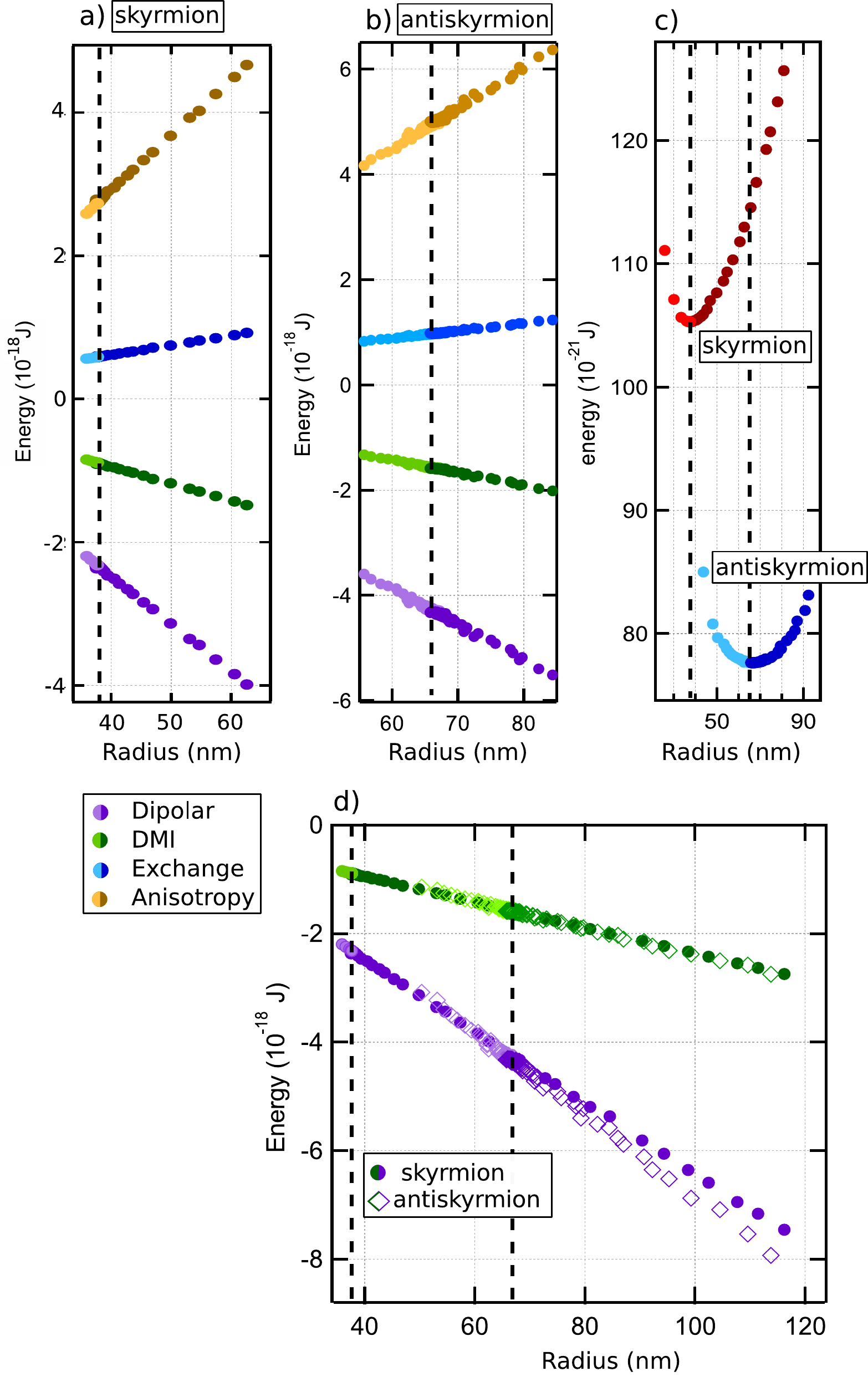}
  \end{center}
  \caption{\label{fig:(R)} Magnetic energies of a skyrmion \textbf{(a)} and an anti-skyrmion \textbf{(b)} as a function of the radius for a given set of magnetic parameters ($M_{\mathrm{s}}= 1.25 \cdot 10^6\,$A/m, $A_{\mathrm{ex}}=16\, $pJ/m,  $K_{\mathrm{eff}}= 2 \cdot 10^5\, $J/m$^3$,  $D= 2.0\, $mJ/m$^2$). \textbf{(c)} Total energy for a skyrmion (red) and an anti-skyrmion (blue) as a function of the radius. \textbf{(d)} Comparison between the DMI (green) and the dipolar interaction (violet) for a skyrmion (dots) and an anti-skyrmion (squares) as a function of the radius. The vertical dotted lines in all panels correspond to the equilibrium radius of Sk and/or ASk}
\end{figure}

Finally, in order to have a numerical confirmation of the Sk and ASk stabilization mechanisms, we studied the energies as a function of the Sk and ASk radius for a given value of $M_{\mathrm{s}}$. We choose $M_{\mathrm{s}} = 1.25\cdot 10^{6}\,$A/m in order to study the regime where the dipolar interactions are important. These simulations were performed starting from two different initial states, respectively with a radius larger and smaller than the equilibrium one. The conjugate gradient minimization algorithm has been used for minimizing the energy, starting from these initial states. During the relaxation towards equilibrium, all the components of the Sk and ASk energies have been tracked as a function of the radius, i.e. for each minimization step we recorded the energies and calculated the radius of the Sk or the ASk [Fig.~\ref{fig:(R)}(a,b)]. Implicitly we suppose here that the configurations follow a physical minimum energy path, which is not necessarily always the case as shown using more complex methods \cite{Dittrich2004,Bessarab2015,Rohart2016,Bessarab2017,Rohart2017}. In order to confirm our results, we compared the obtained energy with the analytical results, as a function of the texture diameter. In both approaches the anisotropy and the DMI energies are proportional to the radius as expected.

Upon diameter increase, anisotropy and exchange energies increase and DMI and dipolar energies decrease, all almost linearly. The balance between these terms is rather subtle as all these energies almost compensate (the absolute value of the total energy is more than one order of magnitude smaller than the absolute value of any of the separate energies).
In Fig.~\ref{fig:(R)}(c) we show that for a given set of magnetic parameters the ASk is more stable than the Sk and it has a bigger radius. It can be understood considering Fig.~\ref{fig:(R)}(d) where the behavior of the DMI and of the dipolar interaction energies are compared as a function of the radius. One can notice that the DMI has the same behavior for the Sk and ASk, unlike the dipolar energy, which upon increasing radius decreases faster for the ASk than for the Sk. This difference is the fundamental reason for the energy difference between the Sk and the ASk.
Even if this difference at equilibrium is not visible in the energy range shown in Fig.~\ref{fig:(R)}(d) it becomes fundamental in the anti-skyrmion/skyrmion energy range Fig.~\ref{fig:(R)}(c).
Indeed the Sk and ASk configurations are determined by the competition between all the magnetic energies and any small variation of one of the energies can imply a strong change of the Sk and ASk energy and radius.

\section{Conclusions}
We have shown that when the dipolar interactions are neglected it is possible to write the ASk energy in a circular symmetric form. The Sk and the ASk in systems with different symmetry but the same strength of magnetic interactions have the same size and stability energy. The presence of dipolar interactions breaks the circular symmetry of the ASk energy. With the support of micromagnetic simulations we have studied the energy and the shape of Sk and ASk as a function of M$_\mathrm{s}$ and explain the role of the dipolar interaction. We can distinguish three different effects. The interaction due to the surface charges does not break the circular symmetry and stabilizes in the same way Sk and ASk. The volume charges depend on the in-plane moment configuration. While the Sk configuration shows homochiral N\'{e}el moment rotation, anti-skyrmions show partly N\'{e}el and partly Bloch rotations. The latter do not produce magnetic charges. The ASk configuration is therefore more stable and the tendency to favor Bloch rotations induces a square shape. Moreover the presence of N\'{e}el rotations with different chirality induces a partial flux closure effect and further increases the ASk stability.
Since both Sk and ASk are stable when all the magnetic energies cancel each other, a small variation of a single parameter like the dipolar energy can have a large influence on the shape and energy of the textures.

\section{Acknowledgements}
We want to express our thanks to Florian Dadoushi and R\'emi Dupouy for the fundamental support in the development of data analysis software and the Agence National de la Recherche (ANR), project ANR-14-CE26-0012 (ULTRASKY), and the Laboratoire d'excellence LANEF (Grant No. ANR-10-LABX-51-01) for financial support.


\begin{thebibliography}{42}
\bibitem[{dis()}]{discret}
\bibinfo{note}{As discretization on a square lattice in simulations may break
  the circular symmetry, we optimized the mesh size so that an isotropic medium
  provides isotropic textures. In the most challenging case, for a skyrmion in
  a 400nm round dot with $M_\mathrm{s} = 1.8\times10^6$~A/m, mesh sizes of 1, 2 and 4 nm
  give respectively 0.99 , 0.97 and 0.94 circularities. The 1 nm mesh size has
  thus been used in all micromagnetic simulations shown in this paper.}
\end{thebibliography}
\end{document}